\documentclass[preprint,amsmath,amssymb,aps,twocolumn,10pt,superscriptaddress]{revtex4-1}

\usepackage{graphicx}
\usepackage{graphics}
\usepackage{epsf}
\usepackage{dcolumn}
\usepackage{comment}
\usepackage{bm}

\newcommand{\angstrom}{\mbox{\normalfont\AA}}

\begin{document}


\title{The influence of Gaussian strain on sublattice selectivity of impurities in graphene}

\author{James A. Lawlor}
 \email{Corresponding author. E-mail: jalawlor@tcd.ie (James Lawlor)}
 \affiliation{School of Physics, Trinity College Dublin, Dublin 2, Ireland.}
\author{Claudia G. Rocha}
 \affiliation{School of Physics, Trinity College Dublin, Dublin 2, Ireland.}
\affiliation{
CRANN, Trinity College Dublin, Dublin 2, Ireland.
}%
\author{Vanessa Torres}
 \affiliation{Instituto de F\'{\i}sica, Universidade Federal Fluminense, Niteroi-RJ, Brazil}
\author{Andrea Latg\'{e}}
 \affiliation{Instituto de F\'{\i}sica, Universidade Federal Fluminense, Niteroi-RJ, Brazil}

\author{Mauro S. Ferreira}
 \affiliation{School of Physics, Trinity College Dublin, Dublin 2, Ireland.}
\affiliation{
CRANN, Trinity College Dublin, Dublin 2, Ireland.
}%

\date{\today}

 \begin{abstract}
 Among the different strategies used to induce the opening of a band gap in graphene, one common practice is through chemical doping.
 While a gap may me opened in this way, disorder-induced scattering is an unwanted side-effect that impacts the electron mobility in the conductive regime of the system.
 However, this undesirable side effect is known to be minimised if dopants interact asymmetrically with the two sublattices of graphene.
 In this work we propose that mechanical strain can be used to introduce such a sublattice asymmetry in the doping process of graphene.
 We argue that a localised out-of-plane deformation applied to a graphene sheet can make one of the graphene sublattices more energetically favourable for impurity adsorption than the other and that this can be controlled by varying the strain parameters.
 Two complementary modelling schemes are used to describe the electronic structure of the flat and deformed graphene sheets: a tight-binding model and density functional theory.
 Our results indicate a novel way to select the doping process of graphene through strain engineering. 

\end{abstract}

\maketitle

\section{Introduction}




Graphene, often described as the ``wonder material'', was once regarded as the ultimate replacement for silicon in nanoelectronic devices. The prospects for such a substitution have gradually faded as a result of poor on/off current ratios evidenced in graphene transistors \cite{jang2013,lee2015,berrada2013}. In fact, the main obstacle inhibiting graphene-based electronics to thrive is enabling a perfect off-state with absolute zero current. It turns out that pristine graphene is a zero gap semiconductor. As a consequence, it can be hard to produce a perfect insulating state \cite{oostinga2007}. Therefore, the search for mechanisms that can efficiently tailor its energy band gap is the only way to pave its usage as a building-block component in digital applications. Mechanical deformation and chemical doping/adsorption are some of the strategies widely employed to engineer a band-gap in the electronic structure of graphene. Both schemes have shown to be effective in tuning the electronic structure but perhaps ultimate levels of control can be achieved by combining these two methods. 

The reason graphene is a zero-gap material lies in its perfect hexagonal structure which can be thought of as two inter-penetrating triangular sublattices, hereafter referred to as A and B. Its low-energy linear dispersion is protected by the symmetry existent between A and B \cite{skomski2014} and breaking this symmetry is the key ingredient for opening a gap. A direct way of tailoring graphene's band gap mechanically is through the application of uniaxial strain \cite{pereira2009tight,ni2008}. Yet, graphene gapless Dirac spectrum can be quite resilient against the application of planar strain. Theoretical predictions have demonstrated that bulk graphene can support elongations up to 20\% without splitting the degeneracy of the bands at the Fermi level \cite{pereira2009tight}. Sublattice symmetry can be better lifted by inducing out-of-plane deformations in graphene structures such as local centrosymmetric bumps \cite{schneider2015,latge2014,settnes2015patched,moldovan2013electronic} and strained nanobubbles \cite{levy2010strain}. According to continuum models in which perturbations are described as effective gauge fields, the local distortion in the graphene structure emulates the presence of a pseudomagnetic field \cite{vozmediano2008} that breaks sublattice symmetry.

As previously mentioned, the electronic structure of graphene can also be tuned via adsorption of impurities or chemical doping. Upon the presence of foreign atoms, distinct mechanisms can dictate the opening of a band gap in graphene. For a random spreading of dopants over the graphene host, disorder is the dominant mechanism that induces the opening of a band gap. Nonetheless,  gap engineering can also be achieved throughout an intriguing ``ordered'' doping state as reported by several experimental contributions \cite{zhao2011visualizing,lv2012nitrogen,zabet2014segregation, wang2012review}. In this state, the majority of dopants manifest a preference for occupying a particular graphene sublattice. One suggestion is that this asymmetric doping is driven by oscillations in the local density of states (LDOS) caused by the impurities themselves \cite{lawlor2014sublattice}. In other words, the impurities are the ones responsible for breaking the sublattice symmetry and making one sublattice energetically more favourable than the other. 

In this work, we propose a way to impel foreign objects to favour a particular sublattice by inducing a Gaussian-shaped deformation on the graphene sheet.
 Gaussian strain, also known as a centrosymmetric Gaussian deformation, have been used to study graphene materials with different extra confinement potentials such as disks, rings, and nanoribbons \cite{wakker2011localization,faria2015fano,carrillo2014gaussian,schneider2015}. 
 Additionally, graphene flakes have been provide natural nanomembranes that can be lift reversibly by the tip of a scanning tunnelling microscope \cite{mashoff2010bistability}.
By centring the Gaussian strain on one particular atomic site (or sublattice), it is known that the sublattice symmetry of the system will be broken yielding pronounced spatial variations in the LDOS, similar to Friedel oscillations \cite{lawlor2013friedel,bacsi2010local}. As a result, we demonstrate that the system binding energy will have a spatial dependency underlined by the shape of the deformed region. This effect dictates the placement of impurities at well-defined positions patterned by the Gaussian distortion. We argue that an additional mechanism to modulate the transport properties of graphene is possible when its sublattice symmetry is broken through the application of mechanical strain. 

\section{Method and calculation details}

To demonstrate how out-of-plane deformations in graphene structures can interfere on the adsorption
process of adatoms, two theoretical methodologies were used: (i) a nearest-neighbour tight-binding (TB) description in conjunction with the patched Green function method \cite{settnes2015patched, settnes2015bubbles} and (ii) density functional theory (DFT) \cite{siesta1,siesta2}. In both methods, a hydrogen atom was the probing impurity that is adsorbed directly on top of a carbon atom. 

Within the TB description the nearest-neighbour overlap integral was $t = -2.7$eV, the adatom is modelled as a single level object with on-site energy of $\epsilon_a = 1.8$eV and the electronic hopping describing the C-H bond is given by $\tau = t$. This parametrization was derived by previous works in which the TB band structure is fitted with those extracted via DFT \cite{robinson2008adsorbate,henwood2007ab,arellano2002interaction}.
Although different parametrisations may alter the numerical calculations, the qualitative results obtained in this work remain consistent.

The energetics associated with the binding of an impurity to a host material can be obtained by a general and mathematically transparent methodology called the Lloyd formalism \cite{lloyd1967wave} in which the change in total electronic energy of the system $\Delta E_i$ is given by

\begin{equation}
  \Delta E_i = \frac{2}{\pi} \int_{\infty}^{E_F} dE \, {\rm Im} \ln \det \left(\hat{I} - \hat{g}(E) \hat{V}_i \right)
\label{eq:deb}
\end{equation}
where $i$ is the index labelling the host sublattice site, $E_F = 0$ is the Fermi energy of the system, $\hat{I}$ is the identity matrix, $\hat{g}(E)$ is the energy ($E$) dependent Green function describing the pristine graphene layer and $\hat{V}_i$ is a local perturbing potential describing the connection between the adatom and the graphene host. Note that the adatom can attach to either sublattice and therefore the perturbing potential must be indexed with $i$.

\begin{figure}[t]
  \centering
    \includegraphics[width=0.5\textwidth]{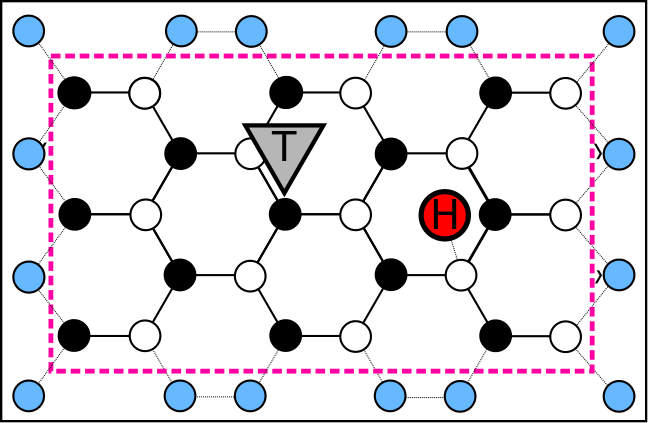}
  \caption{(colour online) (a) Schematic of applying strain to a graphene patch of 3 $\times$ 3 unit cells. The A and B sublattices are represented by black and white circles respectively.
 In this example, Gaussian strain is applied by an idealized atomic tip $T$ positioned right atop of a black site. Shown in blue are the so-called boundary sites which connect the infinite system to the perturbed patch region enclosed by the pink dashed line. The red circle H represents a hydrogen atom adsorbed on a particular host carbon site of the patch.}
  \label{fig:patch_schematic}
\end{figure}

The strain is incorporated by the patched Green function approach where a combination of recursive techniques and bulk lattice Green functions is implemented \cite{power2011electronic}. The Hamiltonian of the central deformed region (cf. Fig. \ref{fig:patch_schematic}) is constructed and then connected to the bulk system via self-energy elements describing the connections between the scattering region and its boundaries. As these elements use the pristine Green functions this ensures the hoppings through the interface are identical to the pristine values. While their values are free to vary within the device region, they return to the pristine values exponentially with the distance from the centre of the strain. This deformation is induced by applying circularly symmetric Gaussian strain centred on a particular atomic site given by the formula

\begin{equation}
\label{eq:strain}
h_j = A e^{-d_j^2/b^2},
\end{equation}
 where $d_j$ is the distance between an atom $j$ and the strain centre and $A$ and $b$ are parameters used to shape the Gaussian deformation; $A$ is associated with its out-of-plane amplitude and $b$ is associated with its circular spreading. The ratio between these (squared) values given by $\alpha = \frac{A^2}{b^2}$ determines the strain intensity generated by the bubble. Here we chose intensities of $\alpha =$ 1\%, 2.5\%, 5\% and 10\%  which are typically found in previous works \cite{de2007charge,faria2013currents,lee2008measurement,cadelano2009nonlinear}. With the whole scattering region deformed according to equation \ref{eq:strain}, we can determine the interatomic distances between neighbouring atoms $m$ and $n$, $l_{mn}$. The resulting change in bond lengths leads to a modification of the nearest neighbour hopping of the tight binding Hamiltonian by
\begin{equation}
\label{eq:t_strain}
t_{mn} = t e^{\beta (\frac{l_{mn}}{l_0} - 1)}
\end{equation}
where $l_0 = 1.42\, \angstrom$ is the length of an unperturbed C-C bond and $\beta = -3.37$ for graphene \cite{pereira2009tight}.
In more details, $l_{mn}$ is the Euclidean distance between atoms $m$ and $n$ after the strain is applied, following equation \ref{eq:strain}. The electronic hoppings $t_{mn}$ are inserted into the Hamiltonian describing the deformed central region from which its associated Green function $\hat{g}$ can be obtained via the patched Green function method. Finally, the binding energy of the system can be calculated through equation \ref{eq:deb}.

\begin{figure}[t]
  \centering
    \includegraphics[width=0.5\textwidth]{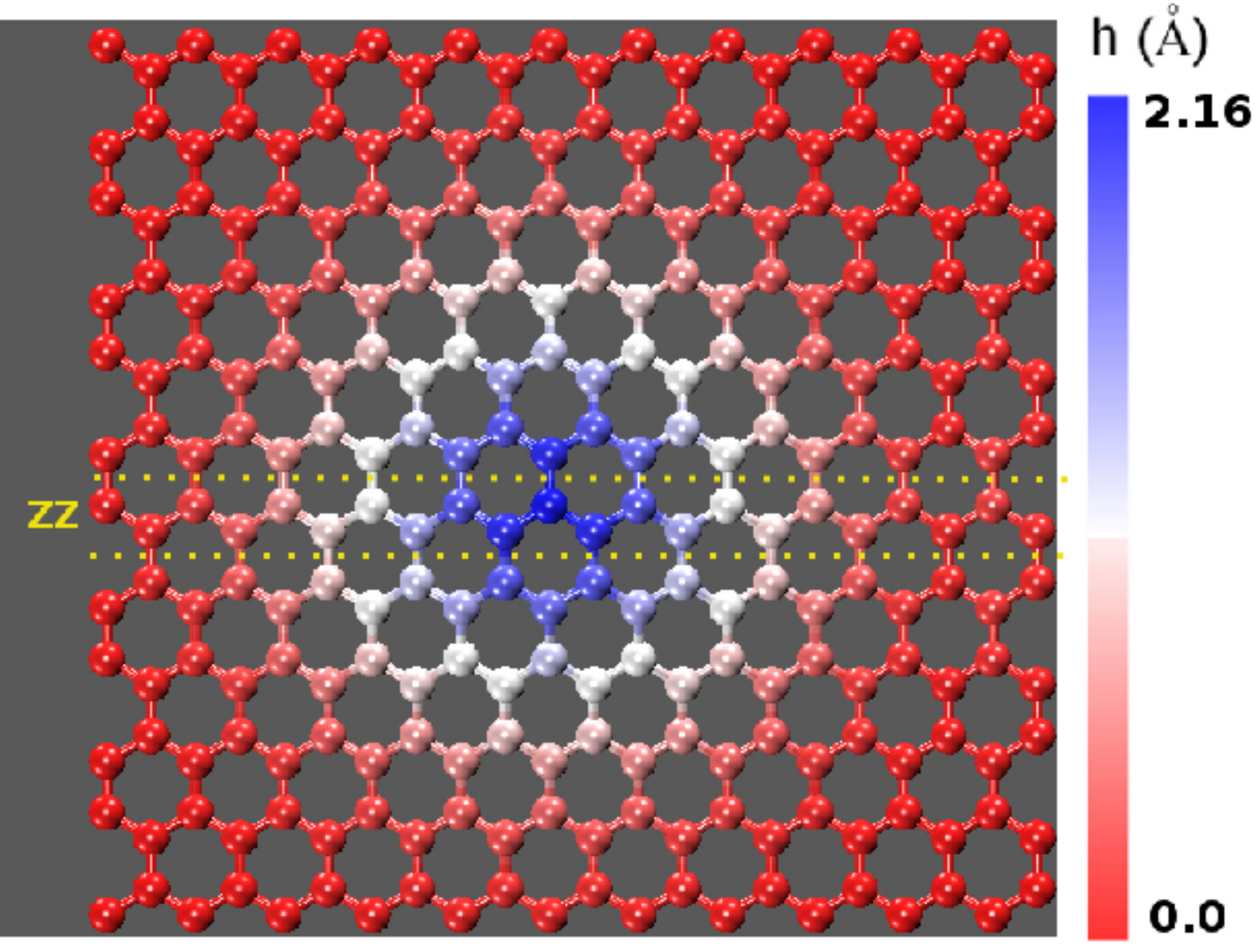}
  \caption{(colour online) Top view of a graphene super cell subjected to a Gaussian deformation of $A = 2.16$ \AA\, and $b = 4.5$ \AA\, ($\alpha = 23\%$). The colour code maps the out-of-plane height ($h$). The centre of the Gaussian is located on the carbon atom enclosed by the yellow circle.
  Dashed yellow lines mark a zigzag segment on which projections on site basis are performed within DFT calculations.}
  \label{fig:dft_relax}
\end{figure}

Besides the TB description DFT calculations were also performed using SIESTA code \cite{siesta1,siesta2} with a generalized gradient approximation (GGA-PBE) \cite{gga} for the 
exchange-correlation potential. In addition,
norm-conserving pseudopotentials with core corrections \cite{pseudopotential} and a split-valence double-$\zeta$ basis (DZP) of pseudoatomic orbitals with an orbital confining energy of 0.05 eV were used.
The energy cutoff was set at 250 Ry. A graphene supercell of approximately 27 $\times$ 26 $\times$ 40 \AA was built in which a Gaussian deformation was induced to alter the atomic positions along $z$-axis. The parameters for the deformation are $A$ = 2.16 \AA, and $b$ = 4.5 \AA\,, leading to $\alpha$ = 23 \% (cf. Figure \ref{fig:dft_relax}).

No structural relaxation was conducted in the deformed sheet and so we guarantee that its Gaussian shape is maintained.
 This deserves special attention as upon relaxation the strain centre moves to the mid-point of the lattice C-C bond length.
 The unrelaxed state can be produced experimentally by maintaining the source of the deformation e.g. an atomic tip.
 Other ways to produce local strain fields include Stone-Wales defects \cite{ma2009stone}, divacancy configurations \cite{lee2006vacancy} and certain substrate arrangements \cite{jean2015topography}.
 While each sublattice is affected equally in the above three cases a sublattice asymmetric effect can be induced by a suitable vacancy or functionalisation on one of the two sublattices or perhaps through a novel combination of these techniques.
 We wish to emphasise that our goal is to show sublattice asymmetry can be produced and enhanced by external factors and need not rely on inter-impurity interactions as has been argued for the large-scale behaviour seen in nitrogen-doped graphene, indeed it should be that any sublattice asymmetric perturbation to the lattice.
 For this reason the sublattice asymmetric Gaussian strain profile employed in this work is not relaxed, structural relaxation was only performed when a Hydrogen atom is adsorbed on top of the carbon atom where the Gaussian deformation is centred. Only atoms in the vicinity of the impurity are allowed to relax. This relaxation was conducted within conjugate gradient algorithm under a 4 $\times$ 4 $\times$ 1 $\Gamma$-centred k-mesh and with a maximum force tolerance of 0.03 eV $\angstrom^{-1}$. The electronic structure of the doped system was subsequently determined in a sampled $\Gamma$-centred Brillouin zone with a grid of 16 $\times$ 16 $\times$ 1. From the electronic structure calculation, we obtained the spatial LDOS distribution within the energy range of [$E_F$ - 0.5 eV, $E_F$].

\section{Results and discussions}

\begin{figure}[t]
  \centering
    \includegraphics[width=0.5\textwidth]{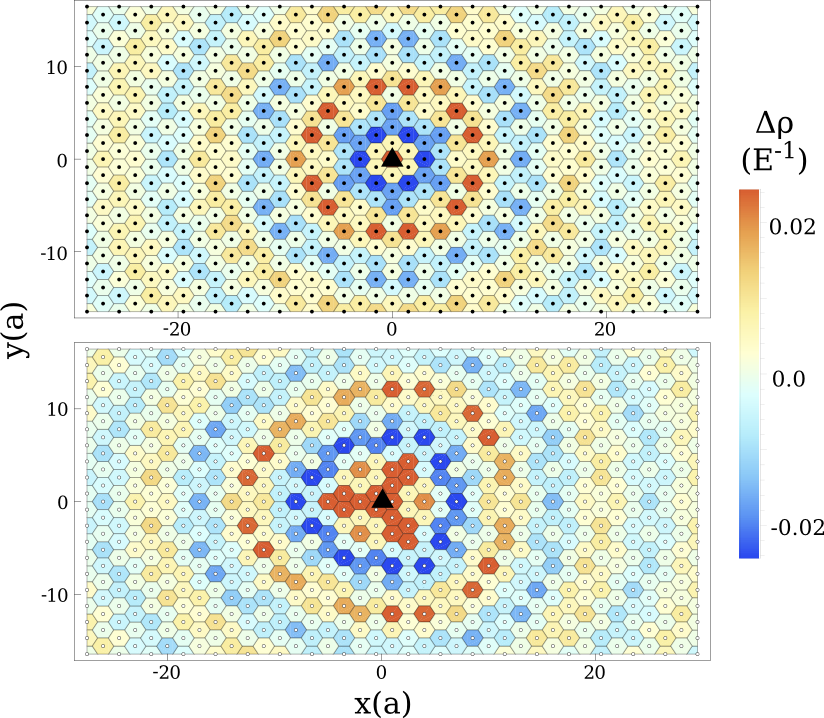}
  \caption{(colour online) Contour plot of LDOS obtained separately for (left panel) A and (right panel) B sublattices for a graphene path deformed with a Gaussian strain of $\alpha = 5\%$. The strain is applied at a carbon atom belonging to sublattice A situated at the origin, indicated on both diagrams by the black triangle. The Fermi energy is set at $E_F = 0.5\,t$.
}
  \label{fig:ldos}
\end{figure}

\begin{figure}[t]
  \centering
    \includegraphics[width=0.45\textwidth]{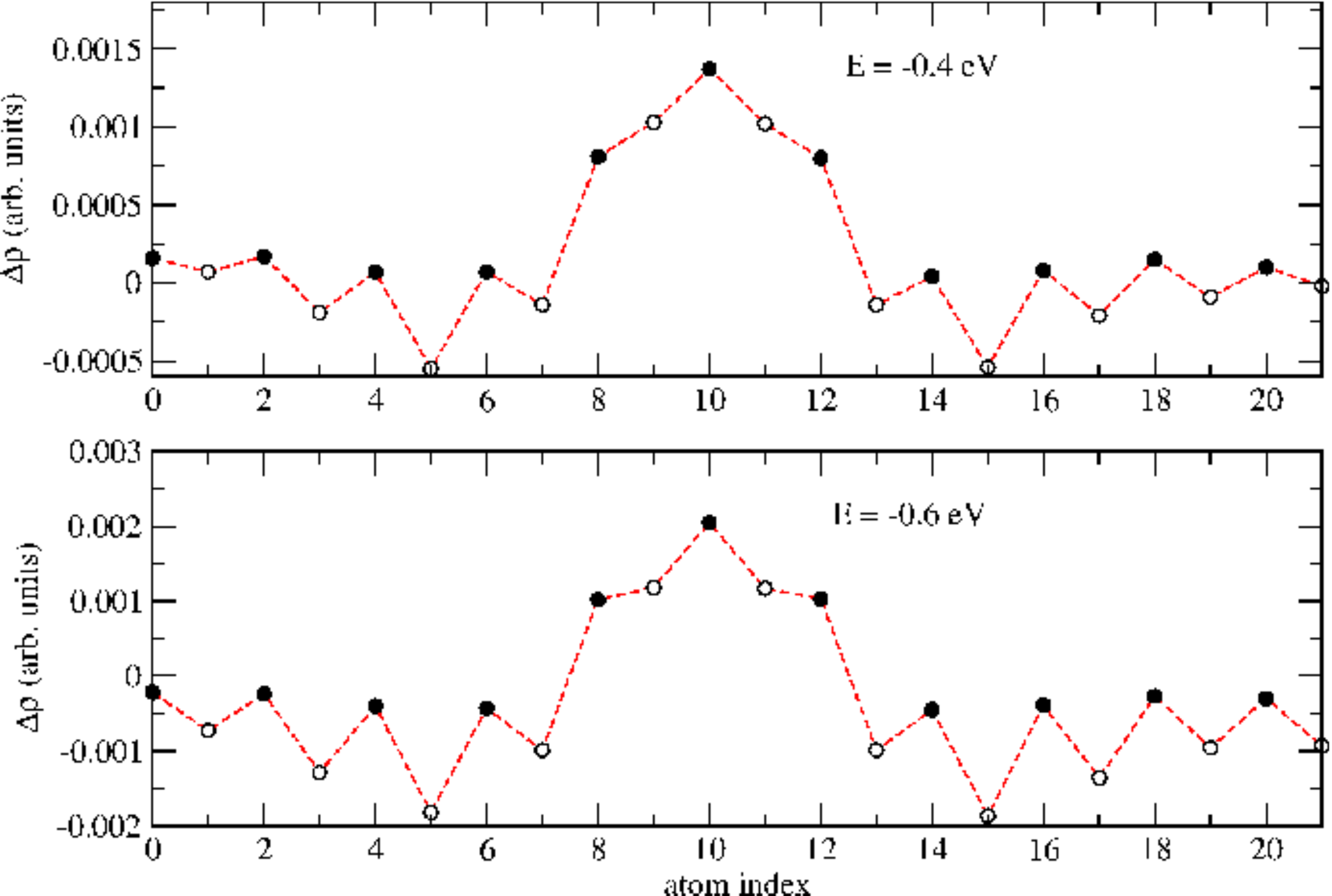}
  \caption{(colour online) LDOS variation ($\Delta \rho$) projected on each carbon site along the horizontal zigzag line highlighted in figure \ref{fig:dft_relax}.
 Solid and hollow points correspond to A and B sublattices respectively.
  This result was obtained within DFT method. The variation was obtained for two fixed energy values: (a) -0.4 eV and (b) -0.6 eV below the Fermi energy of the system.
  The variation is maximum at the site located at the centre of the Gaussian deformation.
}
  \label{fig:ldos_dft}
\end{figure}

To illustrate the distinct signatures exhibited by the two sublattices when pristine graphene undergoes an out-of-plane deformation of $\alpha = 5\%$, we show in figure \ref{fig:ldos} variations in the LDOS per sublattice obtained within the TB picture. This variation is defined as $\Delta \rho = \rho_{bb} - \rho_{fl}$ being $\rho_{bb}$ and $\rho_{fl}$ the LDOS calculated on the graphene bubble and its flat counterpart, respectively. Since TB calculations were done on site-by-site basis, the oscillation profiles are better visualized upon a hexagonal array representation encapsulating each one of the sublattices. In this way, each hexagon encloses a particular carbon site belonging to one of the sublattices (A or B) and its colour maps how intense the LDOS projected on that site is. It is worth mentioning that the system has to be gated for such oscillation patterns to emerge. Additionally, the value $E_F = 0.5\,t$ used to produce figure \ref{fig:ldos} is not easily achievable experimentally and has been used for illustrating the oscillations in a clear way only. At $E_F = 0$, the oscillations disappear as a result of lattice commensurability effects and a continuous $1/D$ decay - $D$ being the spatial distance between any point of the patch and the centre of the bubble - is observed \cite{lawlor2013friedel}.  Furthermore, the LDOS profiles depicted in figure \ref{fig:ldos} resemble Friedel oscillations originated by the adsorption of impurities which in turn break the sublattice symmetry of the system. When strain is applied equally to each sublattice, e.g. at the centre of a hexagon, the resulting LDOS profiles are identical for both sublattices (not shown). This confirms that the centrosymmetric deformation promotes an imbalance in the LDOS distribution with respect to the two inequivalent sublattices \cite{schneider2015}. Similar Friedel-like oscillations are also evidenced in the DFT results. They can be observed in figure \ref{fig:ldos_dft} which depicts $\Delta \rho$ projected on each carbon site located along the zigzag direction highlighted in figure \ref{fig:dft_relax}. 

\begin{figure}[t]
  \centering
    \includegraphics[width=0.42\textwidth]{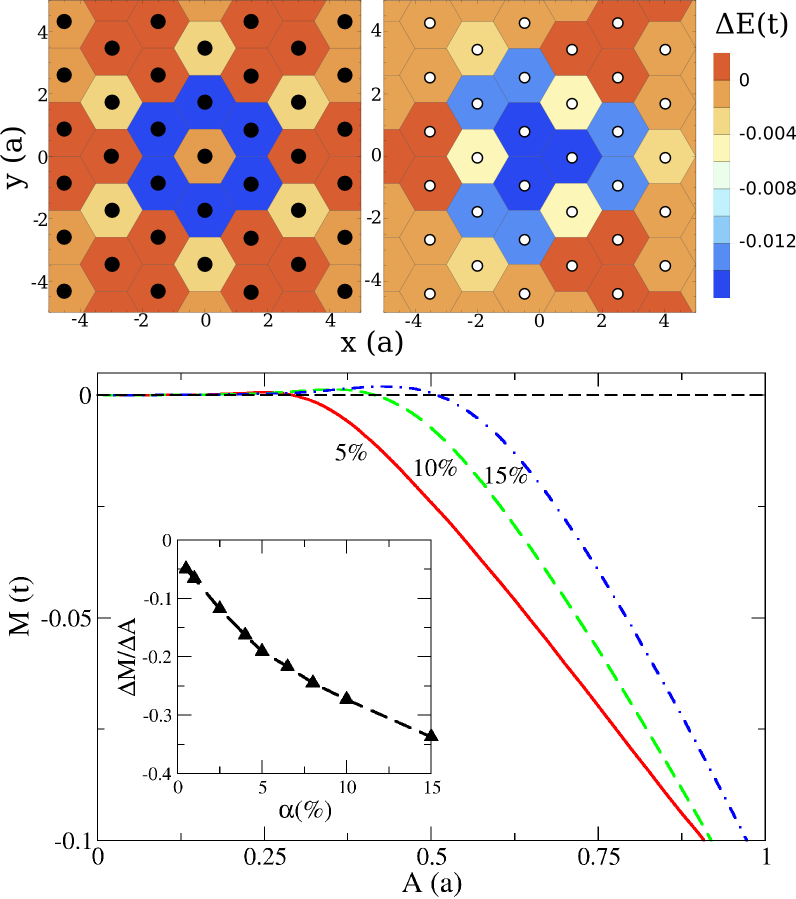}
  \caption{(colour online) (top panels) Contour plots of the spatial variations in $\Delta E$ on the (left panel) A and (right panel) B sublattices for $\alpha = 10\%$, $E_F = 0$, centred at the A sublattice site at the origin. The size of the patch is of 15 $\times$ 15 unit cells.
  (bottom panels) $M$ versus $A$ (strain amplitude) for the same graphene patch (15 $\times$ 15) for distinct values of $\alpha$. The spreading of the bubble is adjusted in such a way that $\alpha$ is fixed at: (red solid) $2.5\%$, (green dash) $5\%$, and (blue dot dash) $10\%$. Inset: gradient of the linear asymptotic behaviour as a function of $\alpha$. Black dashed line serves as a guide to the eyes.}
  \label{fig:sum_vs_strain}
\end{figure}

Similar patterns are observed for the change in electronic energy (cf. eq. \ref{eq:deb}) as a result of adsorbing a H atom on the graphene bubble. 
Figure \ref{fig:sum_vs_strain} shows the short-ranged spatial pattern formed by $\Delta E$ calculated for each adsorption site on the patch subjected to strains of $\alpha = 10\%$ at $E_F = 0$.
The plots are separated according to the sublattices A (black circles) and B (white circles). 
The same hexagonal representation used in the LDOS profiles is used in the energy variation results. One can see that the energy variation profiles for each carbon sublattices (A and B) differ largely, although both of them exhibit a particular pattern determined by the sublattice on which the impurity is placed. Such sublattice disparity would not be evidenced if the path was flat. The shape of the energy profiles is not fixed; it is heavily dependent on the strain intensity and properties of the impurities. In other words, the most susceptible places for binding of impurities can be modulated with respect to $\alpha$.

While it can be seen from the resulting energy profiles that certain atomic sites are more preferable than others, a more descriptive quantity can be obtained by defining a net energy $M$ which accounts for energetic imbalances between the two sublattices.
 This quantity is defined as

\begin{equation}
\label{eq:pseudom}
M = \sum_{j \in \text{B}}\Delta E_j - \sum_{i \in \text{A}} \Delta E_i\,.
\end{equation}
$M$ describes how much more favourable one sublattice is with respect to the other, therefore serving as a useful method to quantify the sublattice asymmetry. Figure \ref{fig:sum_vs_strain} depicts the dependency of $M$ with the amplitude of the Gaussian $A$ for fixed values of $\alpha$. For small amplitudes, $M\approx 0$ indicating that there is no preferable sublattice for the impurity to adsorb. As the out-of-plane deformation increase, $M > 0$ meaning that it is energetically more favourable for the impurity to adsorb on the A sublattice. This scenario is swapped as the level of deformation of the graphene host increases; sublattice B turns to be more favourable permanently where an approximate linear decay of $M$ as a function of $A$ can be observed. In fact, the decay ratio of $M$ depends on the strain intensity as depicted on the bottom inset of figure \ref{fig:sum_vs_strain}. These results confirm that sublattice asymmetry can be achieved via application of Gaussian strain and the most favourable spots for the adsorbants can be tuned with the shape of the bubble.

\begin{figure}[t]
  \center
  \includegraphics[width=0.5\textwidth]{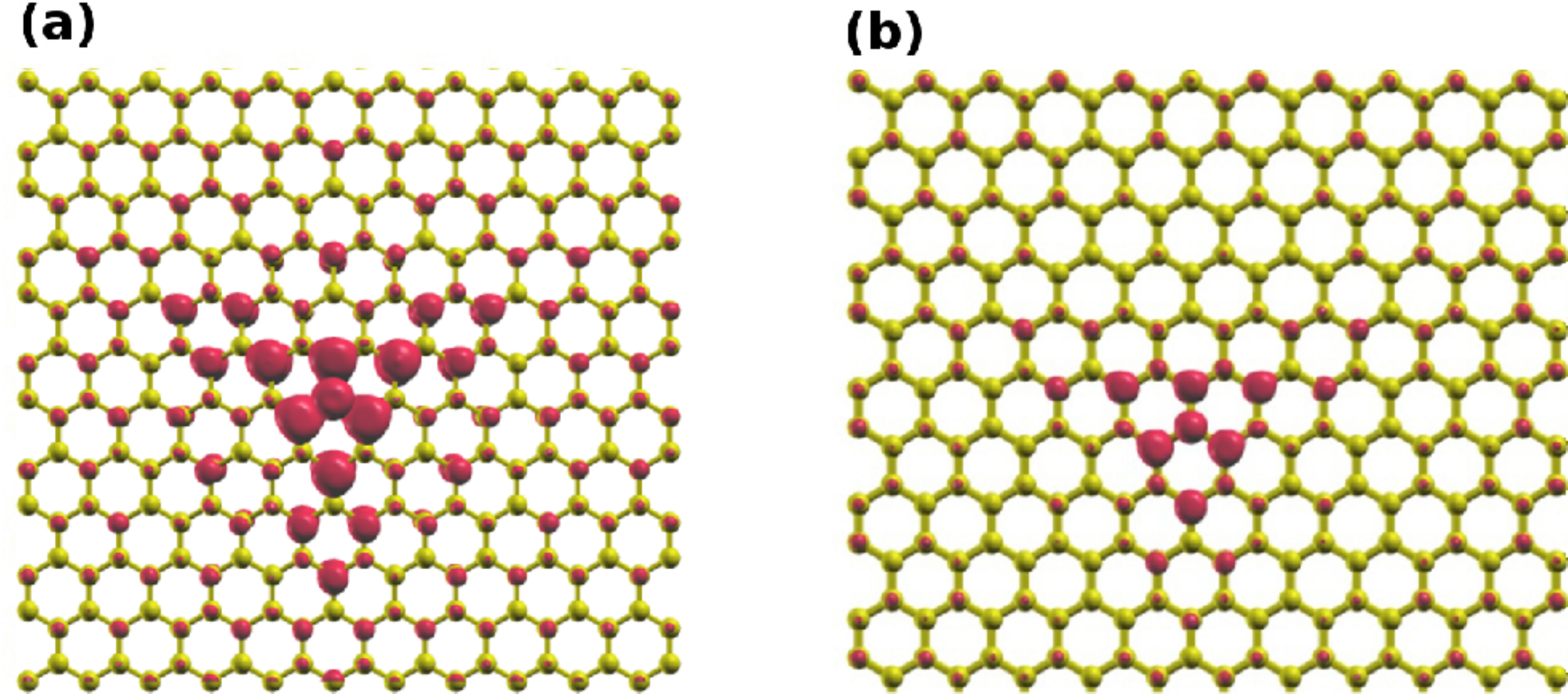}
  \caption{(colour online) Top view of spatial LDOS taken within the energy range [$E_F$ - 0.5 eV, $E_F$] for a doped graphene sheet (a) with and (b) without the Gaussian deformation. The used isosurface value is 0.0007 e/\AA$^3$. The parameters for the deformation are $A$ = 2.16 \AA, and $b$ = 4.5 \AA\,, leading to $\alpha$ = 23 \%.
 The plots were done by using XCRYSDEN visualization tool \cite{xcrysden}.}  \label{fig:dft_ldos}
\end{figure}

To demonstrate the amplified asymmetry effect resulting from the deformation, we show in figure \ref{fig:dft_ldos} the spatial LDOS obtained within DFT for the doped graphene supercell with and without the deformation. A hydrogen atom is adsorbed over the carbon atom in sublattice A upon which the Gaussian deformation is centred.
Hydrogen top-adsorption induces the formation of a sp$^3$-type hybridization with the carbon atom immediately linked to the hydrogen popping slightly off the graphene base \cite{rocha2007modelling}.

This occurs for both deformed and flat graphene platforms.
Therefore, the adsorption itself already breaks the sublattice symmetry; the LDOS exhibits a characteristic threefold spatial dependency which can be observed in figure \ref{fig:dft_ldos}.
It turns out that the deformation induces a much more well-defined LDOS pattern in the considered energy range of [$E_F$ - 0.5, $E_F$] eV. The bubble affects the typical threefold symmetry manifested in the spatial LDOS of top-adsorbed impurities by intensifying the density of states projected on the atoms of sublattice B.
This shows that the Gaussian deformation can be used to modulate the main fingerprint associated to sublattice asymmetry caused by chemical adsorption.
This result also indicates that the out-of-plane distortion establishes a new host environment for subsequent impurities that might adsorb on the bubble opening up new perspectives for band-gap manipulation in graphene structures. 

\section{Conclusions}
In summary, in this work we have demonstrated that mechanical strain in graphene may break the sublattice symmetry that exists in pristine graphene. As a result, the sublattices are no longer equivalent, with one of them becoming more energetically favourable for impurity adsorption. Furthermore, we show that this asymmetry can be controlled with the tuning of the applied strain. This phenomenon is the result of strain-induced oscillations of the local density of states and is somewhat analogous to the impurity-induced Friedel oscillations which have been shown to generate similar features in the spatial distribution of dopants in graphene. While our findings were for hydrogen dopants, it is worth stressing that the results here presented are far more general and are only weakly dependent on the impurity parametrization. As a matter of fact, the determining factor is the adsorption location, which means that similar features should be seen with impurities that have the same bonding geometry as hydrogen, i.e., top-bonded impurities.
 It is well known that realistic graphene samples have a multitude of surface effects such as ripples and corrugations \cite{meyer2007structure,wei2012bending}, and that certain defects can modify the electronic structure and the chemical doping of graphene \cite{wu2012tunable,kim2009effect,fair2013hydrogen}. 
 While the calculations with hydrogen in this work show a slight decrease in the binding energy of hydrogen between pristine and strained graphene, it is a possibility that other impurities and/or strain profiles could produce enhanced chemical doping.
 Our results have shown that variations in local strain have an effect of the sublattice asymmetry throughout the system and that this is driven purely by the asymmetric nature of the applied strain.
 As the external mechanical strain profile increases the competing effects of the ripples on the local electronics will become more diminished. 
Finally, because the transport properties of bi-partite lattices are so sensitive to the spatial distribution of impurities, strain-induced sublattice asymmetry may offer an additional mechanism to engineer the conductance of graphene.

\textit{Author acknowledgements:} JAL, CGR and MSF acknowledge financial support from the Programme for Research in Third Level Institutions (PRTLI). AL and VT acknowledge financial support from CNPq and FAPERJ. MSF also acknowledges financial support from Science Foundation Ireland (Grant No. SFI 11/RFP.1/MTR/3083). The support and computational resources provided by the TCHPC at Trinity College Dublin and the CSC-IT Center for Science in Finland are also acknowledged.
We would also like to thank Stephen Power of Technical University of Denmark for useful discussions.

\bibliographystyle{ieeetr}
\bibliography{bibliography.bib}

\end{document}